\documentclass{raa}
\usepackage{graphicx,times}
\usepackage{multirow}
\usepackage{natbib}
\usepackage{amssymb,amsmath}
\bibpunct{(}{)}{;}{a}{}{,}

\begin{document}

   \title{The photometric system of the One-meter Telescope at Weihai Observatory of Shandong University$^*$
\footnotetext{\small $*$ Supported by the National Natural Science Foundation of China.}
}

 \volnopage{ {\bf 2012} Vol.\ {\bf X} No. {\bf XX}, 000--000}
   \setcounter{page}{1}

   \author{Shao-Ming Hu
   \and Sheng-Hao Han
   \and
   Di-Fu Guo
   \and Jun-Ju Du
    }


   \institute{Shandong Provincial Key Laboratory of Optical Astronomy and Solar-Terrestrial Environment, Institute of Space Sciences, School of Space Science and Physics, Shandong University, Weihai, 264209, China; {\it husm@sdu.edu.cn}\\
\vs \no
   {\small Received     ; accepted    }
}

\abstract{The one-meter telescope at Weihai Observatory of Shandong University is an f/8 Cassegrain telescope. Three sets of filters, including Johnson--Cousins \emph{UBVRI}, Sloan Digital Sky Survey \emph{u$'$g$'$r$'$i$'$z$'$} and Str\"{o}mgren \emph{uvby}, are installed in a dual layer filterwheel. The photometric system and the CCD camera are introduced in this paper, followed by detailed analysis of their performances, and determination of the relevant parameters, including gain, readout noise, dark current and linearity of the CCD camera. In addition, the characteristics of the site astro-climate condition, including typical seeing, clear nights statistics, and average sky brightness were studied systematically based on data gathered from Sep. 2007 to Aug. 2013, and were reported in this work. Photometric calibrations were done using 8 nights Landolt standard star observations, which yielded transformation coefficients, photometry precision and system throughput. The limiting magnitudes were simulated using the derived calibration parameters and classic observation conditions at WHO.
}

   \authorrunning{S. M. Hu et al. }            
   \titlerunning{Photometric system of Weihai 1-m telescope}  
   \maketitle

%
\section{Introduction}           
\label{sect:intro}

The one-meter telescope at Weihai Observatory of Shandong University (hereafter, WHO) was installed in June 2007. It is operated by School of Space Science and Physics, Shandong University, Weihai. This is a Cassegrain telescope made by APM-Telescopes\footnote{http://www.apm-telescopes.de} in Germany. This telescope is similar to the Lulin one-meter telescope (LOT) in Taiwan \citep{kinoshita} and the Tsinghua-NAOC 80-cm telescope (TNT) in Xinglong \citep{huangfang}. Main scientific projects conducted on the WHO 1-m telescope (WHOT) include active galactic nuclei variability \citep{bhatta,hu2014,chen2013,chen2014,guo}, variable stars \citep{yang,dai,likai}, exoplanets, afterglow of Gamma-ray bursts \citep{xudong}, minor planet searching and so on. Two instruments were built and attached to WHOT, including one imaging CCD camera equipped with several sets of filters, and one fiber-fed high resolution Echelle spectrograph (hereafter, HRS). It is essential to know the properties and performance, such as detection limit, throughput, photometry precision and instrument response of the system, which will help the astronomers to prepare observation proposals well (e.g. \citealt{kinoshita,zhou,mao,zhang} ).

A program to characterize the photometric system has been carried out, and the results are reported in this work. The performance of the HRS will be given in another paper. The instruments and control software are briefly introduced in Sect.~\ref{sect:instrument}. The site conditions are reported in Sect.~\ref{sect:site}. Basic characteristics of the CCD camera are shown in Sect.~\ref{sect:ccd}. The photometrical calibration results are described in Sect.~\ref{sect:calibration}. The system performance is given in Sect.~\ref{sect:performance}, and Sect.~\ref{sect:summary} gives a summary.

\section{The observation system and the control software}
\label{sect:instrument}

 WHOT has a `classical' Cassegrain design with a focal ratio of 8 by Lomo Optics. It has a parabolic primary mirror with an effective diameter of 1000\,mm (1020\,mm mechanical diameter), and a hyperbolic secondary mirror with an effective diameter of 360\,mm. Over the entire field of view (FOV), 80\% energy is concentrated with 0.65 arcsec point spread function. Aluminium plus SiO$_2$ is coated to both mirrors. The mount system is an equatorial fork mount with high accuracy fiction servo-drivers. The maximum slew speed can reach to 4 degrees per second in both Right Ascension and Declination directions accelerated by 48 Volts DC motors. The pointing accuracy is 5.4$''$ (RMS) for altitude higher than 20 degrees, and the tracking accuracy can reach to 0.6$''$ (RMS) in 10 minutes blind guiding.

 A back-illuminated PIXIS 2048B CCD camera from the Princeton Instruments Inc.\footnote{http://www.princetoninstruments.com/} is mounted to the Cassegrain focus of WHOT.  Dark current rate is low at a temperature of $-55^{\circ}$C, thanks to the thermoelectric cooling system. With a 2048$\times$2048 imaging array (13.5$\times$13.5 $\mu$m~pixel$^{-1}$), the camera provides a FOV of 12$'\times$12$'$, with a pixel scale of 0.35$''$ per pixel. The specifications of the CCD chip are given in Sect.~\ref{sect:ccd}. A dual layer filterwheel manufactured by American Astronomical Consultants and Equipment Inc. (ACE) is inserted between the telescope flange and the CCD camera. Each layer consists of 8 cells allowing filters with size 50$\times$50\,mm. Standard Johnson--Cousins \emph{UBVRI}, SDSS \emph{u$'$g$'$r$'$i$'$z$'$} and Str\"{o}mgren \emph{uvby} filters (see \citealt{bessell}, and references therein) are available. The time needed for changing a filter is shorter than 1 second, so the speed is very high.

\begin{figure}
   \centering
    \begin{tabular}{rl}
    \begin{minipage}[t]{5.5cm}
    \vspace{0pt}
    \includegraphics[width=5.4cm]{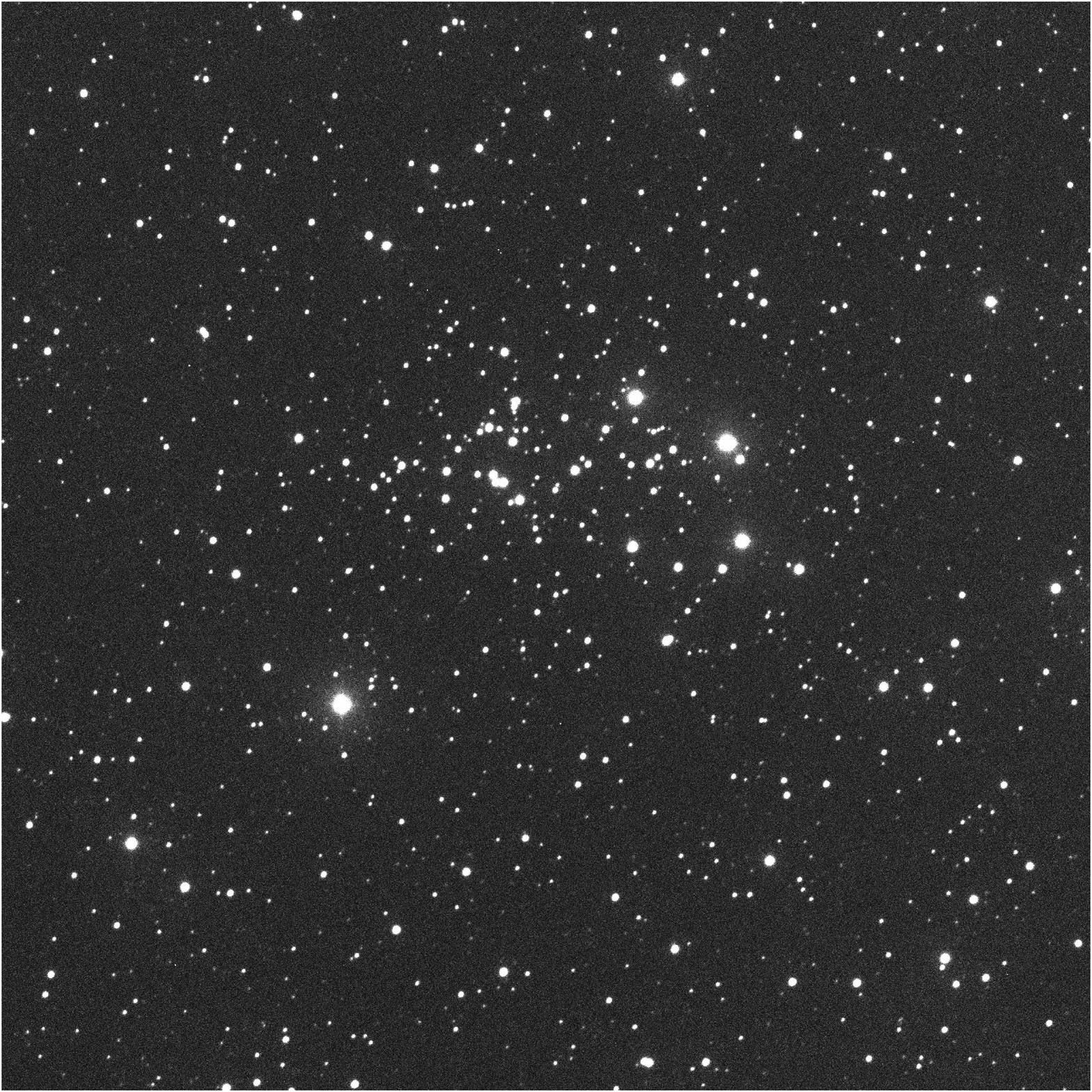}
    \end{minipage}

    \begin{minipage}[t]{7.8cm}
    \vspace{0pt}
    \includegraphics[height=6.0cm, angle=0]{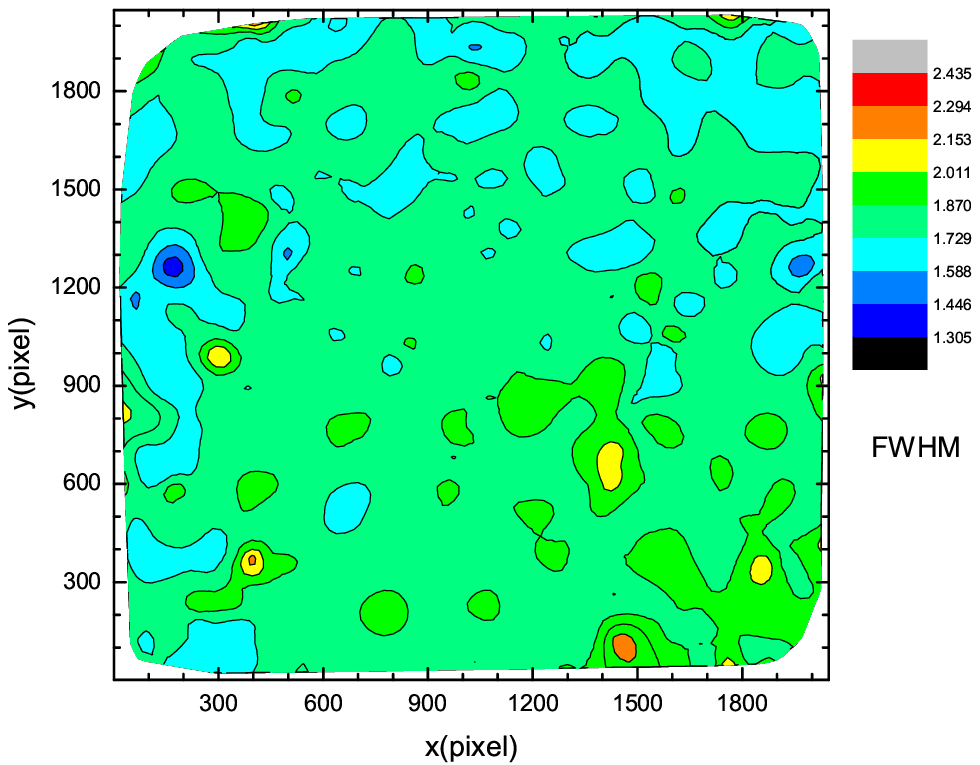}
    \end{minipage}
    \end{tabular}
   \caption{Left: The field of NGC\,7790 taken in $V$ band. Right: The contour of FWHM measured from sources in the field of NGC\,7790.}
   \label{optics}
\end{figure}

 To examine the overall image quality of WHOT, one exposure centered on the field of NGC\,7790 in $V$ band was taken on October 9, 2010 and used to do the full width at half maximum (FWHM) measurement. The image is shown in the left panel of Figure~\ref{optics}. The FWHM of all point sources with signal to noise ratio (SNR) higher than 5 were determined, and its contour is shown in the right panel of Figure~\ref{optics}. This diagram shows that the image quality at focal plane is good and uniform, in the sense that FWHM variation is less than 0.2 arcsec. We note that the lower-right corner has a slightly larger FWHM than the others, which is possibly caused by the mis-alignment of the CCD camera as to optic axis.

 In order to improve the observation efficiency, a software called PCF was developed cooperated with Xinglong station of National Astronomical Observatories. An observation plan file, including for example the target coordinates, filter name, exposure time and the number of repeat observation, should be provided to PCF. PCF will control the telescope to move to the position where the object locates. It will move the dome window and filter wheel to the right position, and control the camera to record the data automatically. It is very efficient and convenient to the astronomers.

\section{Site Conditions}
\label{sect:site}

\begin{figure}
   \centering
    \begin{tabular}{lr}
    \begin{minipage}[b]{7.2cm}
    \includegraphics[width=7cm]{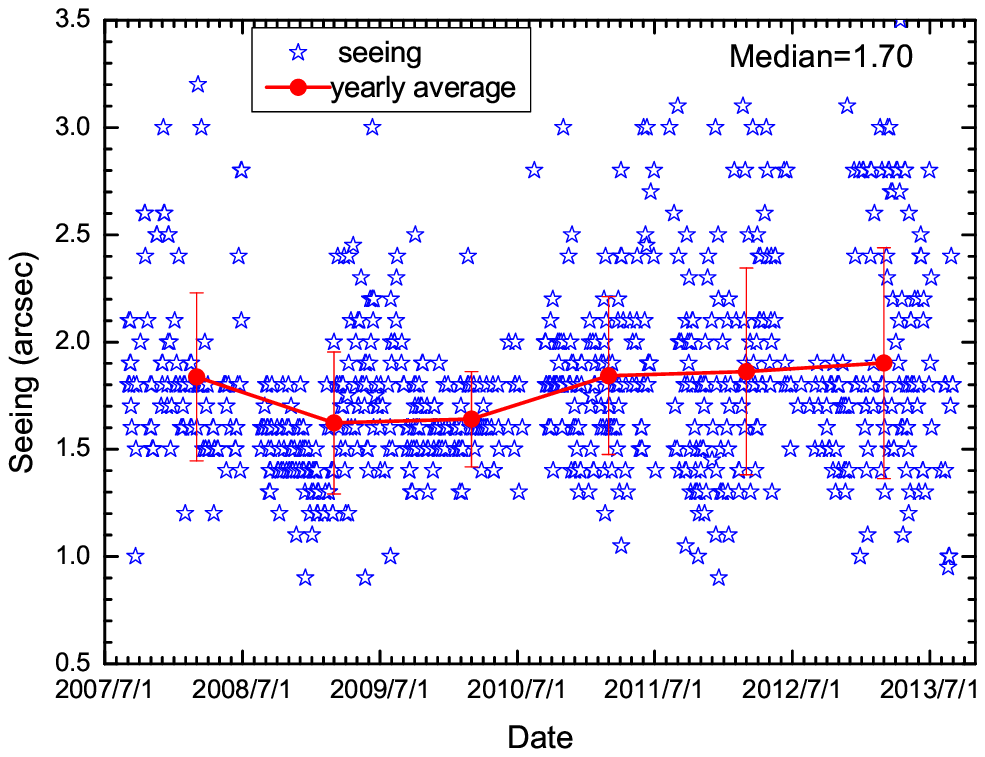}
    \end{minipage}

    \begin{minipage}[b]{7.2cm}
    \includegraphics[width=7cm]{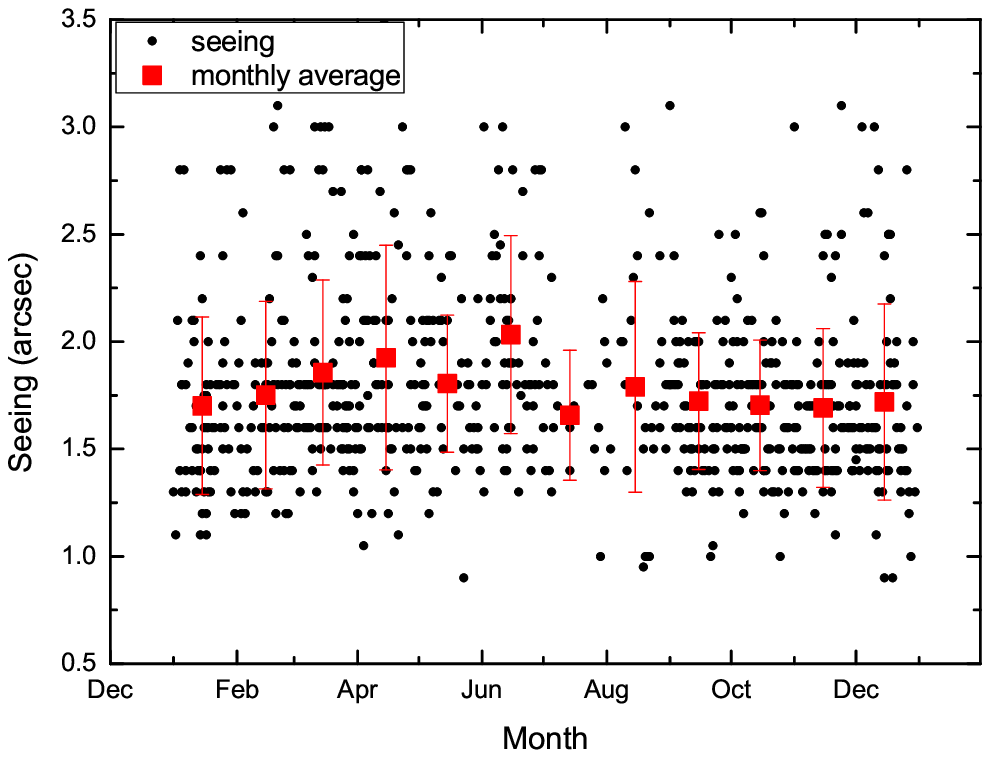}
    \end{minipage}
    \end{tabular}
   \caption{Left: Seeing versus observation date in the past 6 years. Red dots describe the yearly average of the seeing. Right: Seasonal variation of seeing. Red solid squares illustrate the monthly average of the seeing.}
   \label{seeingv}
\end{figure}

\begin{figure}
   \centering
  \includegraphics[width=10cm, angle=0]{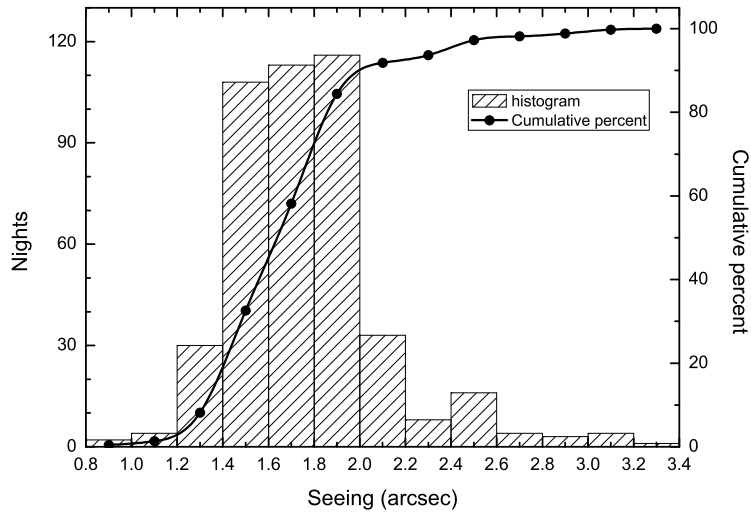}
   \caption{Statistic result of seeing. It is better than 2.0 arcsec on more than 85\% nights.}
   \label{seeings}
\end{figure}

WHOT is located on the top of the Majia Mountain in Weihai (122$^{\circ}$02$'$58.6$''$ E, 37$^{\circ}$32$'$09.3$''$ N), with elevation of about 110 m. Parameters including seeing, number of clear/photometric nights and sky brightness are important to evaluate an astronomical site \citep{zou,yao,zhang}. To evaluate the site seeing conditions of WHO, we have done the statistic analysis using the data recorded in the past 6 years from September 2007 to August 2013. Left panel of Figure~\ref{seeingv} shows the long term seeing value. Because we have no differential image motion monitor (DIMM) to measure the seeing, this value was measured by the FWHM of sources in the image taken at about 14:00 (universal time) on most of observational nights in the past six years. This value includes all the other effects, such as dome seeing, defocus, optical effects and so on \citep{liuli,yao}. The yearly average and standard deviation of seeing in the past six years were calculated and shown by red dots in the left panel of Figure~\ref{seeingv} (the standard deviation is used as error bar). The seeing condition is stable in the past six years. The small variation is probably due to a sampling bias, because the criteria for opening telescope for observations change from year to year. To analyze the seasonal variation of the seeing, we plot all the seeing data versus the date in a year in the right panel of Figure~\ref{seeingv}, and the monthly average is given by red solid squares as well (the standard deviation is used as error bar). It increases slowly from January to April, then it experiences some variations from May to August, finally it almost stabilizes in the rest of  a year. The histogram and cumulative statistic of seeing is shown in Figure~\ref{seeings}. The median of seeing is 1.70$''$. The best seeing can reach to 0.8$''$, and seeing is less than 2.0$''$ on more than 85\% nights.

\begin{figure}
   \centering
    \begin{tabular}{lr}
    \begin{minipage}[b]{7.2cm}
    \includegraphics[width=7cm]{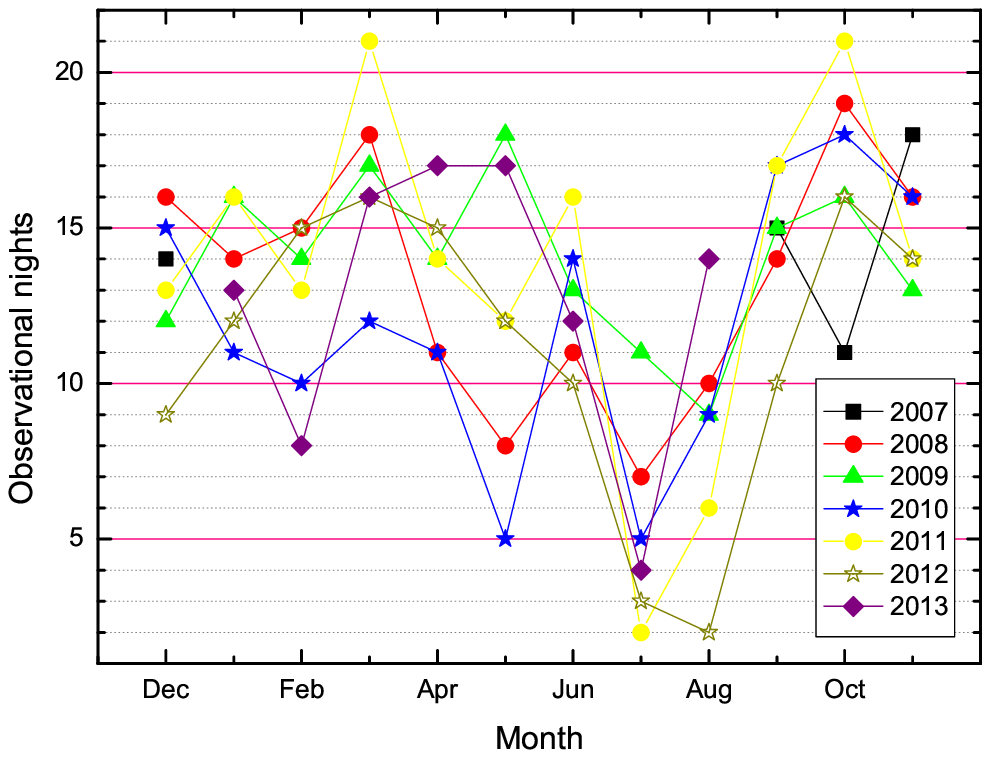}
    \end{minipage}

    \begin{minipage}[b]{7.2cm}
    \includegraphics[width=7cm]{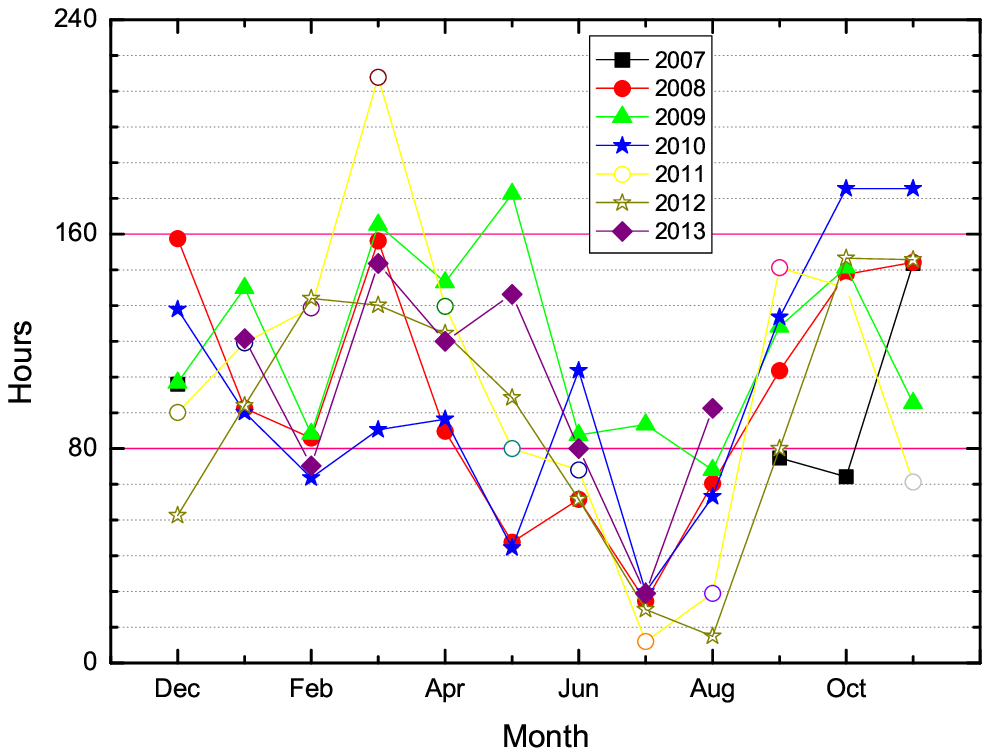}
    \end{minipage}
\end{tabular}
\caption{Monthly observational time in each month during the past six years. Numbers of monthly observational nights and monthly observational hours were given in the left and right panel, respectively.}
\label{obstime}
\end{figure}

Figure~\ref{obstime} illustrates the real observational time, when useful data was obtained by observation log, in each month during the past six years. Left panel of Figure~\ref{obstime} gives the observational nights in every month and right panel of Figure~\ref{obstime} gives the observational hours in each month. We can see that summer has minimum observational time because of raining and high humidity. The other three seasons are almost the same from the plots. Actually it has more clear nights during the winter, but sometimes we have to close the dome because of the strong wind or frozen snow on the dome. The observational nights varies from 134 to 168 in one year during the past six years, and the average is 155 nights. The observational hour is from 1115 to 1427 hours in a year, and the average is 1226 hours. The total clear nights, which means useful data were taken during the whole night, is 113 from September 2011 to August 2013, so the yearly average total clear nights is 56.5 nights.

\begin{figure}
   \centering
  \includegraphics[width=10cm, angle=0]{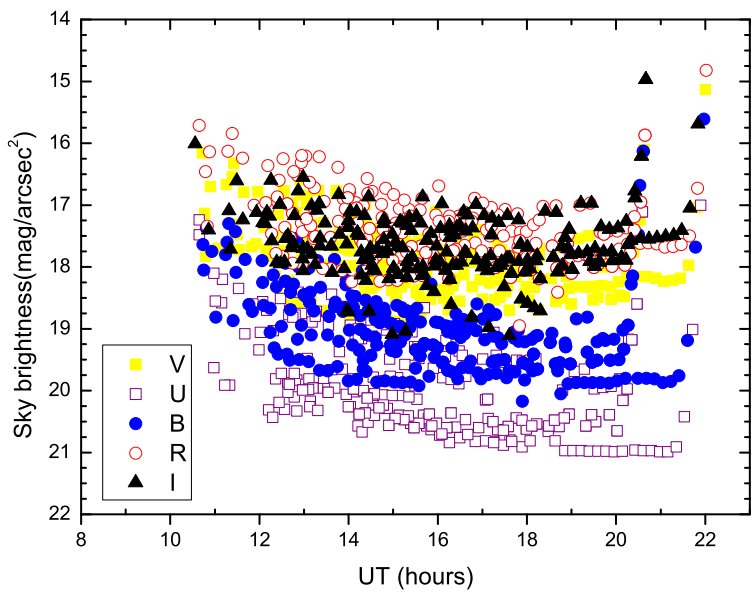}
   \caption{The night sky brightness versus universal time. The purple squares, blue dots, yellow solid squares, red circles and black solid triangles illustrate the night sky brightness in $U$, $B$, $V$, $R$ and $I$ band, respectively.}
   \label{skybrightness}
\end{figure}

 The night sky brightness can be measured using the observations of Landolt standard stars. The night sky count ($Sky_{count}$) can be obtained from the observations during the photometry calibration observations. Then the sky instrumental brightness can be calculated by Equation~(\ref{equsbs}):
\begin{equation}\label{equsbs}
  Sky_{inst}=25-2.5\times log_{10}(Sky_{count}/scale^{2})
\end{equation}
where $Sky_{inst}$ is the instrumental night sky brightness, $scale$ is the angular size of a CCD pixel projected on the sky, i.e. 0.35$''$ per pixel for our configuration. Then the night sky brightness $sky_{x}$, in unit of mag~arcsec$^{-2}$, can be given by Equation~(\ref{equsb}):
\begin{equation}\label{equsb}
  Sky_{x}=Sky_{inst}-ZP_{x}
\end{equation}
where $ZP_{x}$ is the instrument zero point, which can be calculated by transformation equations (see Sect.~\ref{sect:calibration}). $x$ denotes different filter name. Here we did not analyze the effects caused by the difference of the airmass and the direction of the sky area. The \emph{UBVRI} band night sky brightness on 8 nights were plotted against universal time (UT) in Figure~\ref{skybrightness}. An obvious tendency can be seen that the night sky brightness is turning dark before 16:00 (UT, Beijing time is 24:00), then it is stable, finally it brightens just before dawn. Most city lights will be gradually turned off before midnight because they go to sleep. The night sky brightness variation can reach about 2 magnitudes before and after midnight. This tendency strongly supports that the main source of night sky brightness is city artificial light pollution. The median night sky brightness in $U$, $B$, $V$, $R$, and $I$ band is 20.12, 19.14, 18.00, 17.52 and 17.71 mag~arcsec$^{-2}$, respectively. The darkest night sky brightness at WHO in $U$, $B$, $V$, $R$, and $I$  band is 20.99, 20.17, 18.90, 18.95 and 19.11 mag~arcsec$^{-2}$, respectively. It is much brighter than those of the world wide good astronomical sites (see table~6 in \citealt{kinoshita}). It is not a surprise that the night sky brightness in $V$ band is more than 2 mag~arcsec$^{-2}$ brighter than that of Xinglong and Xuyi stations \citep{yao,zhang} because WHO is very close to city. Appropriate research project should be chosen according to site conditions.

\section{CCD characteristics}
\label{sect:ccd}

A PI PIXIS 2048B CCD is used for photometry for WHOT. The peak efficiency of this CCD can reach to 95\%. A new back-illuminated Andor iKon-L DZ936-N CCD camera was ordered for WHOT. It has the same E2V 42-40 chip with PIXIS 2048B, and this camera can be cooled down to $-$80$^{\circ}$C with air cooling. But it is still on the way. So we only describe PIXIS 2048B CCD here.

\begin{table}
\caption{Specifications of PIXIS 2048B CCD camera reported by the manufacturer and our measurements}
\label{tableccd}
\begin{tabular}{c|ccccccc}
\hline
\hline
Attribute  & \multicolumn{7}{|c}{Specifications}\\
\hline
Chip       & \multicolumn{7}{l}{E2V 42$-$40} \\
CCD format & \multicolumn{7}{l}{2048$\times$2048 pixels, pixel size is 13.5 microns, 100\% fill factor.} \\
Image area & \multicolumn{7}{l}{27.6$\times$27.6\,mm$^2$} \\
Cooling    & \multicolumn{7}{l}{Air cooling, down to $-$55$^{\circ}$C} \\
Single pixel full well &  \multicolumn{7}{l}{110, 000\,e$^{-}$} \\
\hline
Dark current(e$^-/$pixel$/$second)  &    &\multicolumn{3}{c|}{ 0.0024  (by manufacturer) @$-$55$^{\circ}$C} & \multicolumn{3}{|c}{ 0.0022  (this work) @$-$55$^{\circ}$C}\\
\hline
\multirow{3}{*}{Non-linearity} & Readout speed & \multicolumn{6}{c}{$<2\%$ (by manufacturer)} \\
\cline{3-8}
                               & 100\,kHz & \multicolumn{3}{c|}{0.67\%(2310-42700\,ADU)}  & \multicolumn{3}{|c}{1.18\%(2310-58950\,ADU)}\\
                               & 2\,MHz   & \multicolumn{3}{c|}{0.49\%(2160-40530\,ADU)}  & \multicolumn{3}{|c}{0.82\%(2160-60200\,ADU)}\\
\hline
\multirow{3}{*}{Readout noise($e^{-}$)} &    & \multicolumn{3}{c|}{Low noise output} & \multicolumn{3}{|c}{High capacity output}\\
\cline{3-8}
                               & 100\,kHz & \multicolumn{3}{c|}{3.67}  & \multicolumn{3}{|c}{7.02}\\
                               & 2\,MHz   & \multicolumn{3}{c|}{15.35} & \multicolumn{3}{|c}{36.23}\\
\hline
\multirow{3}{*}{Gain($e^{-}$/ADU)} &         & Low(1)& Median(2) & High(3)  & Low(1) & Median(2) & High(3)  \\
                      &  100\,kHz & 3.26  & 1.65      &   0.81   & 14.14  &    6.81   & 3.33  \\
                      &  2\,MHz   & 3.59  &  1.83     &   0.97   & 15.58  &    7.41   & 3.88  \\
\hline
\multirow{3}{*}{Readout time(seconds) per frame} &         & Bin mode& 1$\times$1 & 2$\times$2  & 4$\times$4 & 8$\times$8 & 16$\times$16  \\
\cline{3-8}
                                &  100\,kHz &         & 36.45      &  9.521   & 2.595  &    0.738   & 0.288  \\
                                &  2\,MHz   &         & 2.265      &  0.956   & 0.458  &    0.249   & 0.154  \\

\hline
\hline
\end{tabular}

\end{table}

\begin{figure}
   \centering
    \begin{tabular}{lr}
    \begin{minipage}[b]{7.2cm}
    \includegraphics[width=7cm]{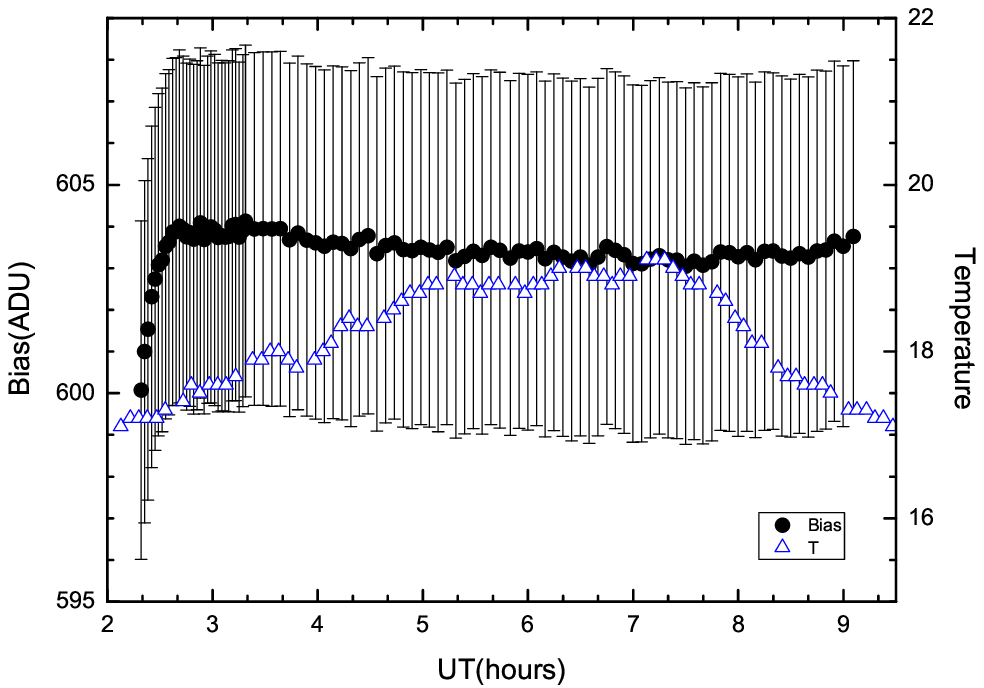}
    \end{minipage}

    \begin{minipage}[b]{7.2cm}
    \includegraphics[width=7cm]{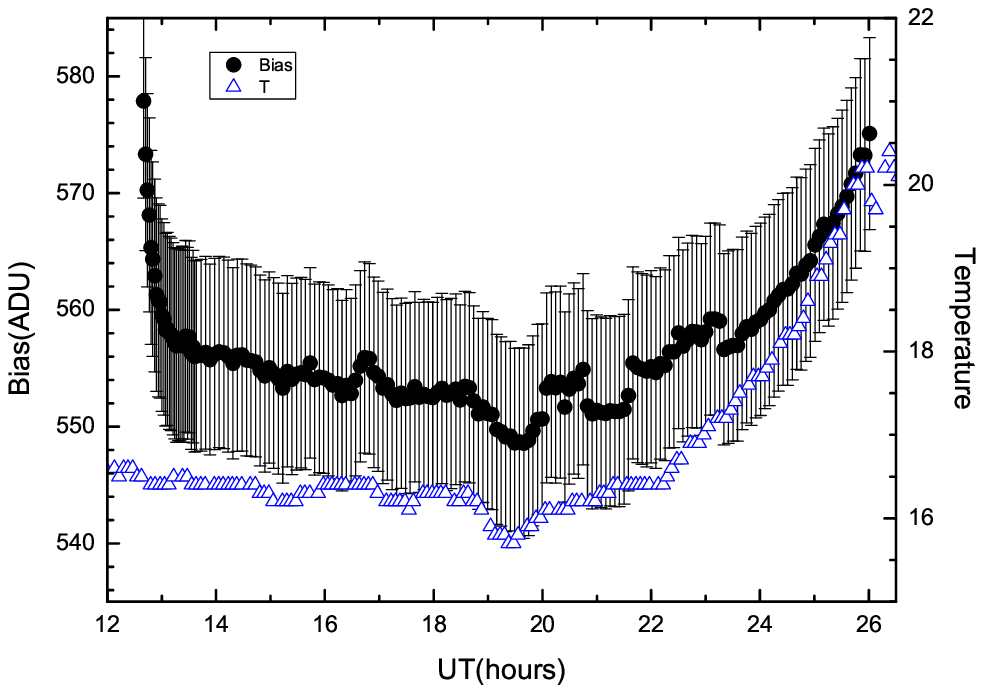}
    \end{minipage}
\end{tabular}
\caption{Means bias level and standard deviation (black), along with ambient temperature (blue) of the camera at the time bias exposure was taken. Data shown in the left panel were in slow readout mode, while that in the right panel in fast readout mode.}
\label{bias}
\end{figure}

The stability of the CCD bias level is a non-negligible effect for high precision photometry. To understand the bias variation well, we monitored the bias level from October 12 to 13, 2011 for more than 6 and 13 hours continuously for slow and fast readout working mode, respectively. One bias image was taken every 2 to 5 minutes, and the average as well as the standard deviation for the center 400$\times$400 pixels were calculated for every image. The environment temperature was recorded by the temperature sensor installed in the weather station. The mean bias level and the temperature against the UT are shown in Figure~\ref{bias}. The result for slow readout working mode is shown in the left panel, while the right panel shows the bias level for the fast readout mode. The triangles illustrate the variation of ambient temperature. The error bar in the plot denotes the standard deviation of the bias. At the beginning of monitoring, the bias changed strongly, because it was not long enough after the CCD dewar was cooled down. This is the reason why we need to cool down the CCD at least 2 hours before scientific images are taken. Usually the CCD camera is kept cooling and stabilizing at $-55^{\circ}C$ all day except the thunderstorm season for the safety of the CCD camera. The bias is stable for slow readout working mode, but it is variable when we set the camera as fast readout mode, and it changed about 25 ADU counts during 13 hours. Its variation seems related to the variation of the ambient temperature. So we need to be careful for high precision photometry when we use the fast readout working mode.

\begin{figure}
   \centering
  \includegraphics[width=10cm, angle=0]{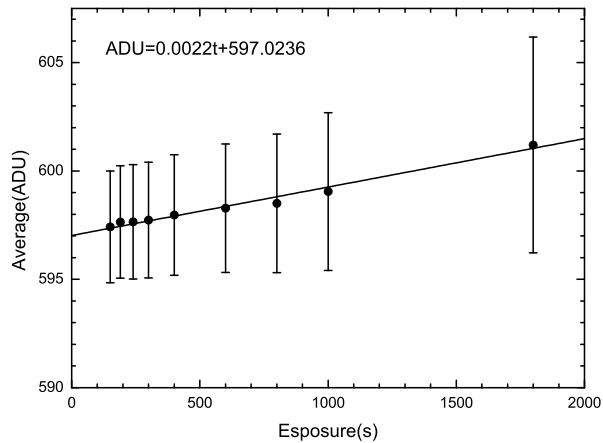}
   \caption{The average ADU count versus the exposure time of dark images. Solid line is the best linear fit between the average ADU count and the exposure time.}
   \label{dark}
\end{figure}

We did the dark current measurement on 2010 July 24. Dark images were taken with exposure times from 150 to 1800\,s, then the average value and the standard deviation for the center 400$\times$400 pixels were calculated for each image. As shown in Figure~\ref{dark}, the average value and exposure time maintain a well-defined linear relationship, whose slope represents the dark current rate. The measured dark current is 0.0022$\pm$0.0001 e$^-/$pixel$/$second (@$-$55$^{\circ}$C), which is very close to the value given by PI. The dark current is very low, so dark correction is unnecessary for usual short exposure photometry observations.

\begin{figure}
   \centering
    \begin{tabular}{lr}
    \begin{minipage}[b]{7.2cm}
    \includegraphics[width=7cm]{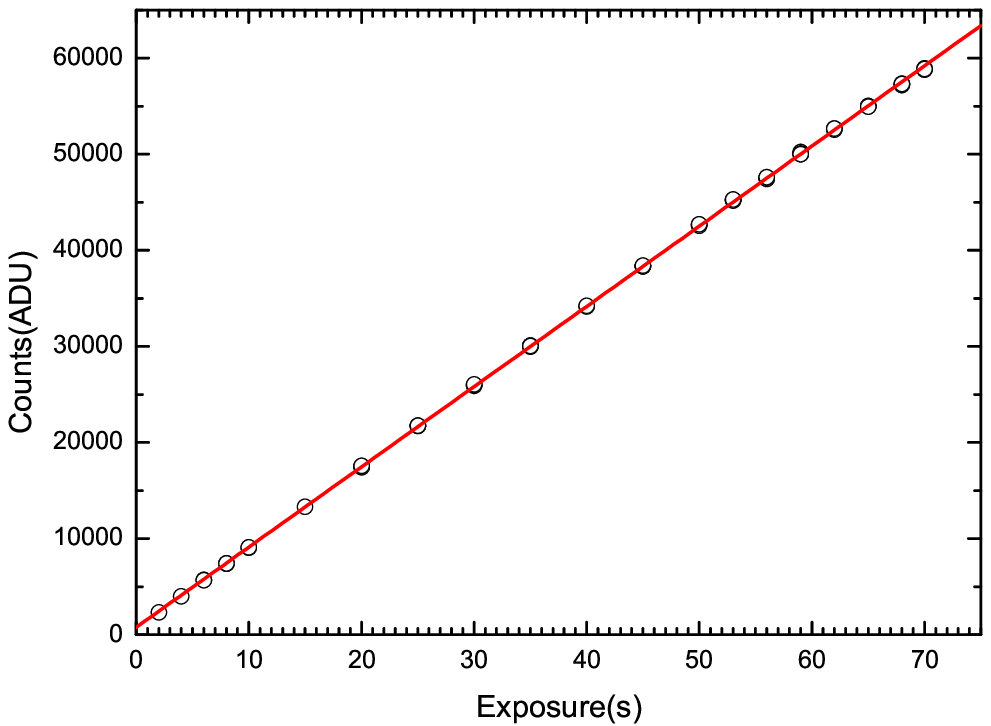}
    \end{minipage}

    \begin{minipage}[b]{7.2cm}
    \includegraphics[width=7cm]{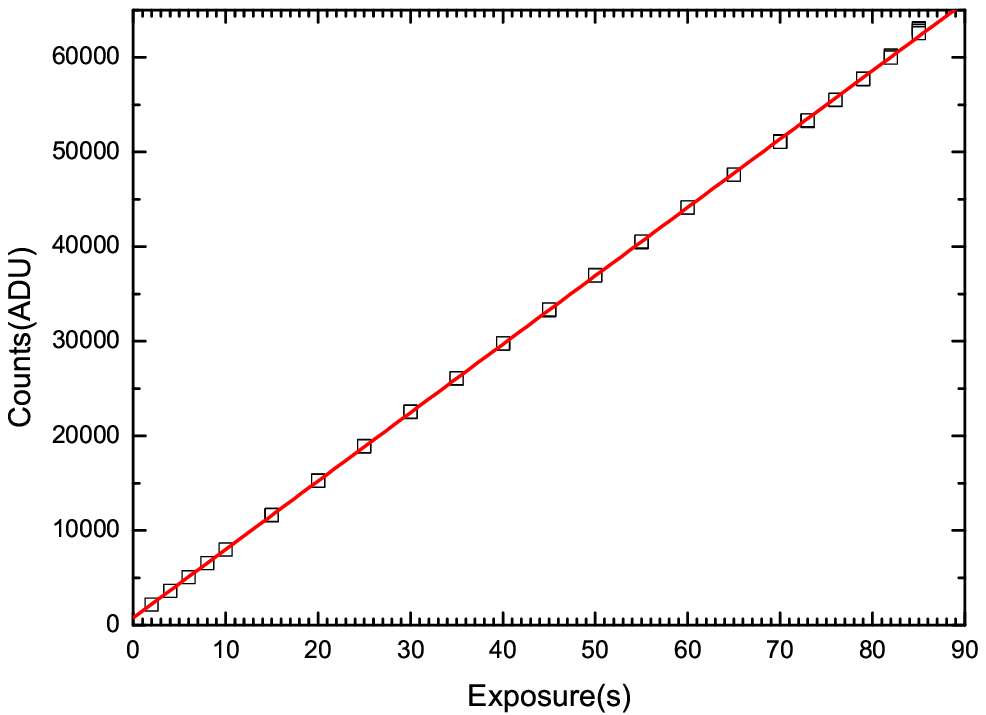}
    \end{minipage}
\end{tabular}
\caption{Linearity of the CCD. The left and right panel illustrate the CCD linearity for slow readout and fast readout speed working mode, respectively. The red lines show the linear least-squares fitting.}
\label{linerity}
\end{figure}

Linearity is quite important to high precision photometry, so the linearity of the CCD response was measured using a series of unfiltered flat-field images with median ADU values from $\sim$ 2000 to $\sim$ 60000. The test was done in a room without light coming in. A fluorescent light was used to cast the room ceiling indirectly, and the camera was covered by several pieces of plain white paper, then the camera was placed upright to take flatfield images. Tests were only done with settings of low noise output and median gain, which is the mostly used setting in the observations. The exposure time varies from 2 to 70\,s for slow readout mode and varies from 2 to 85\,s for fast readout mode. Four images were taken for each exposure time. The average analog digital unit (ADU) was calculated by the center 400$\times$400 pixels for each image. The relationship between the exposure time and the average ADU count for slow readout and fast readout mode is shown in the left and right panel of Figure~\ref{linerity}, respectively. The solid red line in Figure~\ref{linerity} is the linear least-squares fitting of measured data points. The nonlinearity can be calculated by the following equation:
\begin{equation}\label{nonlinearity}
  Nonlinearity(\%)=\frac{MaxPositiveDev+MaxNegativeDev}{MaxSignal} \times 100
\end{equation}
where $MaxPositiveDev$ and $MaxNegativeDev$ is the measured maximum positive and negative deviation from the linear fitting, respectively. $MaxSignal$ is the measured maximum of the ADU counts. By measurement and calculation, the nonlinearity is 0.67\% and 1.18\% at 42700 and 58950 ADU, respectively with slow readout speed, while it is 0.49\% and 0.82\% at 40530 and 60200 ADU, respectively with fast readout speed. Other specifications of the detector, such as gain, readout noise, readout time are listed in Table~\ref{tableccd}.

\section{Photometrical calibration}
\label{sect:calibration}

In order to compare between various telescopes/instruments, photometric calibration must be done, i.e., to convert instrumental magnitudes into standard systems. The transformation coefficients for Johnson-Coursins UBVRI system are measured and reported in the current work. The transformation equations are defined as follows:
\begin{equation}\label{equu}
  U_{inst}=U_{std}+Z_{U}+K_{U}'X+C_{U}(U-B)_{std}
\end{equation}
\begin{equation}\label{equu}
  B_{inst}=B_{std}+Z_{B}+K_{B}'X+C_{B}(B-V)_{std}
\end{equation}
\begin{equation}\label{equu}
  V_{inst}=V_{std}+Z_{V}+K_{V}'X+C_{V}(B-V)_{std}
\end{equation}
\begin{equation}\label{equu}
  R_{inst}=R_{std}+Z_{R}+K_{R}'X+C_{R}(V-R)_{std}
\end{equation}
\begin{equation}\label{equu}
  I_{inst}=I_{std}+Z_{I}+K_{I}'X+C_{I}(V-I)_{std}
\end{equation}
where $U_{inst}$, $B_{inst}$, $V_{inst}$, $R_{inst}$, $I_{inst}$ are the instrumental magnitudes, $U_{std}$, $B_{std}$, $V_{std}$, $R_{std}$, $I_{std}$ are the standard magnitudes, $Z_{U}$, $Z_{B}$, $Z_{V}$, $Z_{R}$, $Z_{I}$ are the zero points of the transformation equations, $K_{U}'$, $K_{B}'$, $K_{V}'$, $K_{R}'$, $K_{I}'$ are the first-order extinction coefficients, $C_{U}$, $C_{B}$, $C_{V}$, $C_{R}$, $C_{I}$ are the color terms of the transformation equations, and $X$ denotes the airmass.

\begin{figure}
   \centering
  \includegraphics[width=10cm, angle=0]{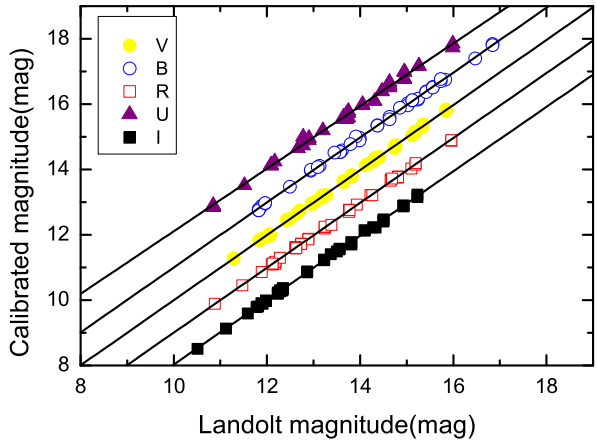}
   \caption{Standard magnitude (Landolt) versus calibrated magnitude from derived transformation equations for \emph{UBVRI} band. The observations was done on 2008 October 24. The $U$, $B$, $R$, $I$ calibrated magnitudes were shifted 2, 1, $-$1, $-$2 magnitudes, respectively for clarity. Solid lines are the best linear fit of the data points.}
   \label{calibration}
\end{figure}

To derive the above parameters, we observed a lot of Landolt standards \citep{landolt} with a wide range of color on photometric nights covering a wide range of airmass. Many standards were observed on 8 nights from 2008 to 2013. All the images were reduced by standard photometry steps, and the same aperture radius of 7$''$ as \citet{landolt} was used for photometry, then the instrumental magnitudes and the airmass were obtained. The standard magnitudes and colors are taken from \citet{landolt}, so all the parameters could be solved by linear regression. The second-order extinction coefficients were found to be small, so we ignored them in this paper. Figure~\ref{calibration} shows the relationship between the Landolt standard magnitude and the shifted calibrated magnitude using derived transformation coefficients on 2008 October 24. It is shown that the slope of all the fitted lines are quite close to 1.0, so the the calibration is quite good and no systematic errors are recognized. All the solved transformation coefficients for these 8 photometric nights are listed in Table~\ref{tabextinction}. The observation dates are listed in the first column in Table~\ref{tabextinction}. The filter names, zero points ($Z$), first order extinction coefficients ($K'$), color terms, the numbers of standards observed on that night, the numbers of observations per night (N$_o$), median night sky brightness on that night, average system throughput on that night, range of the airmass (minimum$\sim$maximum) and color (minimum$\sim$maximum) are following. The average of the first-order extinction coefficient is 0.778, 0.537, 0.331, 0.241, 0.158 mag/airmass for band $U$, $B$, $V$, $R$, and $I$, respectively. We are not surprised that they are larger than those of Xinglong \citep{huangfang}, Lulin and other best sites in the world (see table~4 and references in \citealt{kinoshita}) because of the low elevation of WHO. They are similar with those at Xuyi \citep{zhang}. The color term of each band is close to the results from TNT and LOT. The mean values of each band are listed in Table~\ref{tabextinction} as well. The byproducts of photometry calibration are night sky brightness and the system throughput. The night sky brightness was given in Sect.~\ref{sect:site}, and the system throughput will be illustrated in Sect.~\ref{sect:performance}.

\begin{table}
\bc
\caption{All the solved transformation coefficients and related parameters.\label{tabextinction}}
\setlength{\tabcolsep}{1pt}
\small
 \begin{tabular}{ccccccccccc}
  \hline\noalign{\smallskip}
Date& Filter & $Z$ & $K'$  & $C$ & N$_s$ & N$_o$ & B$_{sky}$ & Throughput & Airmass & Color \\
  \hline\noalign{\smallskip}
  2008-10-24 & U  &  4.826$\pm$0.096   &  0.648$\pm$0.005   &   $-$0.019$\pm$0.002  &    20  &   21  &   19.00  &     5.9  &  1.2$\sim$2.3  &  $-$1.13$\sim$1.19 \\
  2008-11-29 & U  &  4.476$\pm$0.035   &  0.844$\pm$0.001   &   $-$0.175$\pm$0.001  &     5  &   22  &   19.96  &     6.8  &  1.2$\sim$2.0  &  $-$0.98$\sim$0.14 \\
  2010-03-10 & U  &  3.688$\pm$0.058   &  1.039$\pm$0.002   &   $-$0.163$\pm$0.001  &    17  &   16  &   20.29  &    4.35  &  1.1$\sim$1.8  &  $-$1.11$\sim$1.31 \\
  2010-10-09 & U  &  4.956$\pm$0.097   &  0.650$\pm$0.003   &   $-$0.090$\pm$0.002  &     9  &   16  &   18.90  &     5.1  &  1.1$\sim$2.2  &  $-$1.19$\sim$2.23 \\
  2011-03-04 & U  &  5.180$\pm$0.167   &  0.810$\pm$0.003   &   $-$0.209$\pm$0.001  &    21  &   28  &   19.29  &     4.2  &  1.1$\sim$1.8  &  $-$1.11$\sim$1.31 \\
  2012-02-18 & U  &  3.728$\pm$0.051   &  0.881$\pm$0.005   &   $-$0.232$\pm$0.001  &    17  &    9  &   20.44  &    9.95  &  1.1$\sim$1.9  &  $-$1.11$\sim$1.21 \\
  2013-04-11 & U  &  4.180$\pm$0.125   &  0.814$\pm$0.002   &   $-$0.194$\pm$0.001  &    12  &   24  &   19.77  &    10.9  &  1.3$\sim$2.5  &  $-$1.12$\sim$1.32 \\
  2013-08-31 & U  &  4.385$\pm$0.515   &  0.538$\pm$0.003   &   $-$0.379$\pm$0.001  &     8  &   26  &   19.63  &     7.1  &  1.1$\sim$2.5  &  $-$1.19$\sim$2.31 \\

   \hline
        mean &    &  4.427$\pm$0.548   &  0.778$\pm$0.159   &   $-$0.183$\pm$0.105      &                &   & & & &\\
   \hline
  2008-10-24 & B  &  2.841$\pm$0.043   &  0.420$\pm$0.001   &   $-$0.132$\pm$0.001  &    25  &   18  &   18.96  &    13.9  &  1.2$\sim$2.4  &  $-$0.24$\sim$1.42 \\
  2008-11-29 & B  &  2.791$\pm$0.016   &  0.442$\pm$0.001   &   $-$0.177$\pm$0.001  &     5  &   16  &   19.69  &    13.5  &  1.2$\sim$2.0  &  $-$0.24$\sim$0.68 \\
  2010-03-10 & B  &  2.176$\pm$0.051   &  0.623$\pm$0.001   &   $-$0.164$\pm$0.001  &    21  &   18  &   19.35  &    26.3  &  1.1$\sim$1.8  &  $-$0.27$\sim$1.64 \\
  2010-10-09 & B  &  2.915$\pm$0.072   &  0.460$\pm$0.003   &   $-$0.124$\pm$0.001  &    11  &   17  &   18.60  &    12.4  &  1.1$\sim$2.4  &  $-$0.32$\sim$2.00 \\
  2011-03-04 & B  &  3.218$\pm$0.074   &  0.596$\pm$0.008   &   $-$0.169$\pm$0.001  &    22  &   11  &   18.75  &     9.8  &  1.1$\sim$1.8  &  $-$0.27$\sim$1.91 \\
  2012-02-18 & B  &  2.115$\pm$0.063   &  0.679$\pm$0.001   &   $-$0.082$\pm$0.001  &    24  &   40  &   19.70  &    27.0  &  1.1$\sim$2.0  &  $-$0.27$\sim$1.91 \\
  2013-04-11 & B  &  2.651$\pm$0.065   &  0.461$\pm$0.001   &   $-$0.094$\pm$0.001  &    20  &   27  &   18.84  &    17.2  &  1.3$\sim$2.5  &  $-$0.29$\sim$1.91 \\
  2013-08-31 & B  &  2.443$\pm$0.112   &  0.616$\pm$0.001   &   $-$0.165$\pm$0.001  &    13  &   54  &   19.26  &    20.7  &  1.1$\sim$2.8  &  $-$0.32$\sim$2.00 \\
   \hline
       mean  &    &  2.644$\pm$0.378   &  0.537$\pm$0.101   &   $-$0.138$\pm$0.036  &                &   & & & &\\
   \hline
  2008-10-24 & V  &  2.815$\pm$0.032   &  0.271$\pm$0.001   &    0.043$\pm$0.002  &    24  &   19  &   17.70  &    22.3  &  1.2$\sim$2.5  &  $-$0.24$\sim$1.42 \\
  2008-11-29 & V  &  2.640$\pm$0.020   &  0.283$\pm$0.001   &    0.063$\pm$0.001  &     5  &   16  &   18.71  &    26.0  &  1.2$\sim$2.0  &  $-$0.24$\sim$0.68 \\
  2010-03-10 & V  &  2.151$\pm$0.040   &  0.407$\pm$0.001   &    0.069$\pm$0.001  &    20  &   20  &   18.02  &    38.8  &  1.1$\sim$1.9  &  $-$0.27$\sim$1.91 \\
  2010-10-09 & V  &  2.923$\pm$0.086   &  0.257$\pm$0.004   &    0.121$\pm$0.001  &    11  &   17  &   17.40  &    19.7  &  1.1$\sim$2.5  &  $-$0.32$\sim$2.00 \\
  2011-03-04 & V  &  3.058$\pm$0.041   &  0.372$\pm$0.009   &    0.072$\pm$0.001  &    21  &   11  &   17.49  &    15.5  &  1.1$\sim$1.8  &  $-$0.27$\sim$1.64 \\
  2012-02-18 & V  &  2.377$\pm$0.046   &  0.364$\pm$0.001   &    0.081$\pm$0.001  &    24  &   40  &   18.16  &    28.8  &  1.1$\sim$2.0  &  $-$0.27$\sim$1.91 \\
  2013-04-11 & V  &  2.800$\pm$0.053   &  0.351$\pm$0.001   &    0.062$\pm$0.001  &    20  &   26  &   17.76  &    22.2  &  1.3$\sim$2.5  &  $-$0.29$\sim$1.91 \\
  2013-08-31 & V  &  2.591$\pm$0.100   &  0.342$\pm$0.001   &    0.003$\pm$0.001  &    13  &   54  &   18.32  &    28.9  &  1.1$\sim$2.8  &  $-$0.32$\sim$2.00 \\
   \hline
        mean &    &  2.670$\pm$0.296   &0.331$\pm$0.054     &    0.064$\pm$0.033  &                &   & & & &\\
   \hline
  2008-10-24 & R  &  2.505$\pm$0.031   &  0.176$\pm$0.001   &    0.076$\pm$0.003  &    25  &   19  &   17.07  &    26.9  &  1.2$\sim$2.5  &  $-$0.12$\sim$0.93 \\
  2008-11-29 & R  &  2.552$\pm$0.010   &  0.222$\pm$0.001   &    0.063$\pm$0.001  &     4  &   16  &   18.06  &    25.6  &  1.2$\sim$2.1  &  $-$0.12$\sim$0.31 \\
  2010-03-10 & R  &  2.000$\pm$0.064   &  0.263$\pm$0.001   &    0.194$\pm$0.001  &     9  &    9  &   17.52  &    38.8  &  1.1$\sim$1.9  &  $-$0.14$\sim$1.22 \\
  2010-10-09 & R  &  2.509$\pm$0.039   &  0.219$\pm$0.003   &    0.183$\pm$0.001  &    11  &   16  &   17.04  &    25.8  &  1.1$\sim$2.5  &  $-$0.15$\sim$1.17 \\
  2011-03-04 & R  &  2.736$\pm$0.034   &  0.280$\pm$0.007   &    0.126$\pm$0.002  &    20  &   10  &   16.79  &    19.7  &  1.1$\sim$1.8  &  $-$0.14$\sim$1.22 \\
  2012-02-18 & R  &  1.965$\pm$0.027   &  0.298$\pm$0.002   &    0.079$\pm$0.001  &    17  &    9  &   17.67  &    33.3  &  1.1$\sim$2.0  &  $-$0.14$\sim$1.53 \\
  2013-04-11 & R  &  2.470$\pm$0.038   &  0.219$\pm$0.001   &    0.095$\pm$0.001  &    13  &   24  &   17.26  &    27.4  &  1.3$\sim$2.6  &  $-$0.13$\sim$1.29 \\
  2013-08-31 & R  &  2.402$\pm$0.089   &  0.250$\pm$0.001   &    0.104$\pm$0.001  &    13  &   53  &   17.81  &    30.4  &  1.1$\sim$2.9  &  $-$0.15$\sim$1.18 \\
   \hline
        mean &    &  2.392$\pm$0.271   &  0.241$\pm$0.039   &    0.115$\pm$0.049  &                &   & & & &\\
   \hline
  2008-10-24 & I  &  3.017$\pm$0.037   &  0.121$\pm$0.001   &   $-$0.037$\pm$0.001  &    24  &   18  &   17.17  &    18.1  &  1.2$\sim$2.5  &  $-$0.26$\sim$1.84 \\
  2008-11-29 & I  &  3.239$\pm$0.038   &  0.128$\pm$0.001   &   $-$0.044$\pm$0.001  &     3  &   10  &   18.06  &    14.3  &  1.2$\sim$2.1  &  $-$0.26$\sim$0.63 \\
  2010-03-10 & I  &  2.512$\pm$0.025   &  0.203$\pm$0.001   &   $-$0.054$\pm$0.001  &    19  &   18  &   17.67  &    29.0  &  1.1$\sim$1.9  &  $-$0.30$\sim$2.79 \\
  2010-10-09 & I  &  2.935$\pm$0.039   &  0.190$\pm$0.002   &   $-$0.016$\pm$0.001  &    11  &   16  &   17.23  &    19.0  &  1.1$\sim$2.6  &  $-$0.33$\sim$2.27 \\
  2011-03-04 & I  &  3.031$\pm$0.114   &  0.287$\pm$0.005   &   $-$0.058$\pm$0.001  &    16  &   11  &   14.63  &    17.6  &  1.1$\sim$1.8  &  $-$0.30$\sim$2.79 \\
  2012-02-18 & I  &  2.563$\pm$0.032   &  0.165$\pm$0.001   &   $-$0.058$\pm$0.001  &    23  &   35  &   17.76  &    31.8  &  1.1$\sim$2.0  &  $-$0.30$\sim$3.48 \\
  2013-04-11 & I  &  2.957$\pm$0.055   &  0.136$\pm$0.001   &   $-$0.085$\pm$0.001  &    13  &   24  &   17.30  &    20.0  &  1.3$\sim$2.6  &  $-$0.29$\sim$2.97 \\
  2013-08-31 & I  &  2.867$\pm$0.083   &  0.246$\pm$0.001   &   $-$0.050$\pm$0.001  &    13  &   51  &   17.88  &    21.9  &  1.1$\sim$2.7  &  $-$0.33$\sim$2.40 \\
  \hline
        mean &    &  2.890$\pm$0.243   &  0.185$\pm$0.059   &   $-$0.050$\pm$0.020  &                &   & & & &\\

  \noalign{\smallskip}\hline
\end{tabular}
\ec
\end{table}

\section{System performance}
\label{sect:performance}

\subsection{System Throughput}

 The total throughput of the entire optical system can be estimated by observing a series of standard stars. The throughput involves the telescope optics, the filter response, the detector quantum efficiency, and the atmospheric transmission. Following \citet{kinoshita}, measurements of the system throughput of WHOT in each band were done on photometric nights. As shown in Col.~9 of Table~\ref{tabextinction}, the throughput is not constant, which may due to hardware change, dust, atmosphere extinction variation and so on. The total throughput on 2011 March 4 was quite low, because the optic instruments were not cleaned for almost two years. The median throughput of WHOT is 6.4\%, 15.6\%, 24.2\%, 27.2\% and 19.5\% in \emph{U}, \emph{B}, \emph{V}, \emph{R} and \emph{I} band, respectively, and the peak throughput is 10.9\%, 27.0\%, 38.8\%, 38.8\% and 31.8\% in \emph{U}, \emph{B}, \emph{V}, \emph{R} and \emph{I} band, respectively.

\subsection{Limiting magnitude and photometry precision}

\begin{figure}
   \centering
  \includegraphics[width=10cm, angle=0]{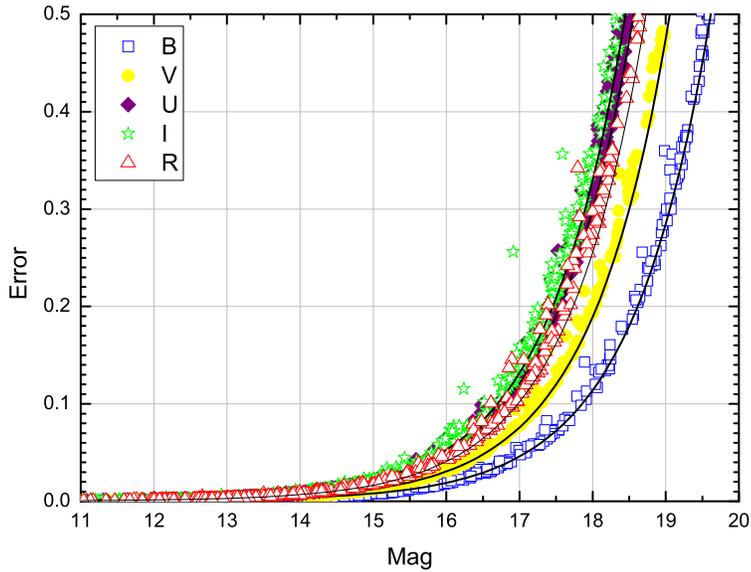}
   \caption{Photometry error versus the brightness of sources. Purple solid diamonds, blue squares, yellow solid circles, red triangles and green stars illustrate the measurement from M67 fields for $U$, $B$, $V$, $R$ and $I$ band, respectively. Solid lines show the simulation result.}
   \label{magerr}
\end{figure}

Equation~(\ref{equsn}) can be used to calculate the signal to noise ratio of one object observed by CCD camera \citep{howell}:
\begin{equation}\label{equsn}
  SNR=\frac{N_{*}}{\sqrt{N_{*}+n_{pix}(N_{S}+N_D+N_R^2)}}
\end{equation}
where $N_*$ is the total number of photons collected by CCD from the object. It can be calculated by the flux density corresponding to zero magnitude of a specific filter \citep{bessell1979}, if we know the throughput of our telescope and the atmosphere extinction. $N_S$ is the total number of photons in each pixel from the sky background, which can be obtained when the night sky brightness is known. $N_D$ is the ADU dark current per pixel per second, and $N_R$ is the readout noise of the CCD camera. Both of them are the specifications of the CCD camera. $n_{pix}$ is the pixel area considered for SNR calculation, which is related to the seeing of the observation. The photometry error can be given from the signal to noise ratio by Equation~(\ref{equerr}) \citep{howell}:
\begin{equation}\label{equerr}
  \delta=1.0857/SNR
\end{equation}
where $\delta$ is the photometry error for the object. So the photometry error can be simulated according to Equations~(\ref{equsn}-\ref{equerr}). To check the reliability of the simulation, we reduced the 60\,s exposure observations of M67 in $U$, $B$ and $V$ band on 2012 February 18 and in $R$ and $I$ band on 2013 April 11. All the sources with SNR higher than 2 were extracted, then their brightness and photometry error were obtained by standard photometry procedure, and the calibration was done using M67 standards given by Dr. Arne Henden\footnote{http://binaries.boulder.swri.edu/binaries/fields/m67ids.txt}. The relationship between the brightness and the photometry error was shown in Figure~\ref{magerr}. The simulation curves of photometry error versus brightness in five bands, which were simulated using the same night sky brightness, atmosphere extinction coefficients, throughput, exposure time and airmass as the observations, were also plotted by solid lines in Figure~\ref{magerr}. We can see that the simulation fits the observations very well. So the simulation is reliable.

To evaluate the observation ability of WHOT, the simulation was done with 300\,s exposure at zenith, using the average extinction coefficients, the median night sky brightness, the median seeing value (considering radius of 2.5 times of seeing area as SNR calculation) and the peak efficiency of WHOT. Under the same observation conditions as the simulation was done, the limiting magnitude with SNR of 100 and 300\,s exposure is 15.7, 16.7, 16.2, 16.1 and 15.9 mag for $U$, $B$, $V$, $R$ and $I$ band, respectively. The limiting magnitude is much brighter than that of TNT \citep{huangfang} telescope and Xinglong 85-cm \citep{zhou} telescope because of the brighter night sky brightness and bigger atmosphere extinction at WHO. So we have to focus on some projects with bright targets for WHOT. But we should point out that we did not use the smallest extinction coefficients and the darkest night sky brightness at WHO to simulate the limiting magnitude, and the limiting magnitude will be at least half magnitude deeper than what we reported here if we use them.

\begin{figure}
   \centering
    \begin{tabular}{lr}
    \begin{minipage}[b]{7.2cm}
    \includegraphics[width=7cm]{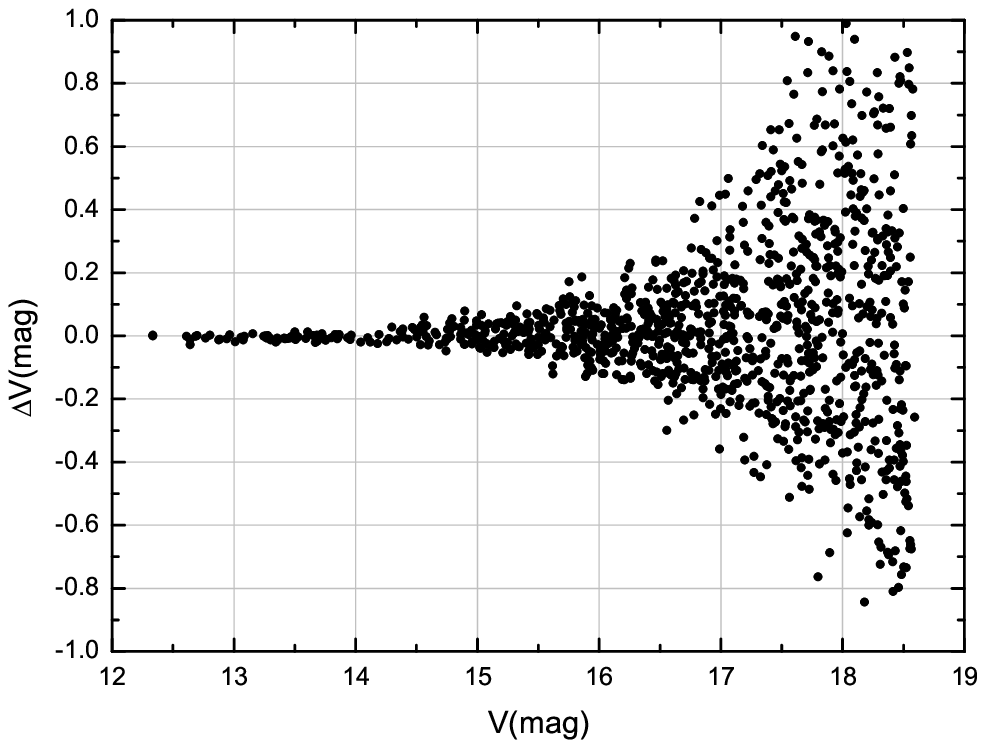}
    \end{minipage}

    \begin{minipage}[b]{7.2cm}
    \includegraphics[width=7cm]{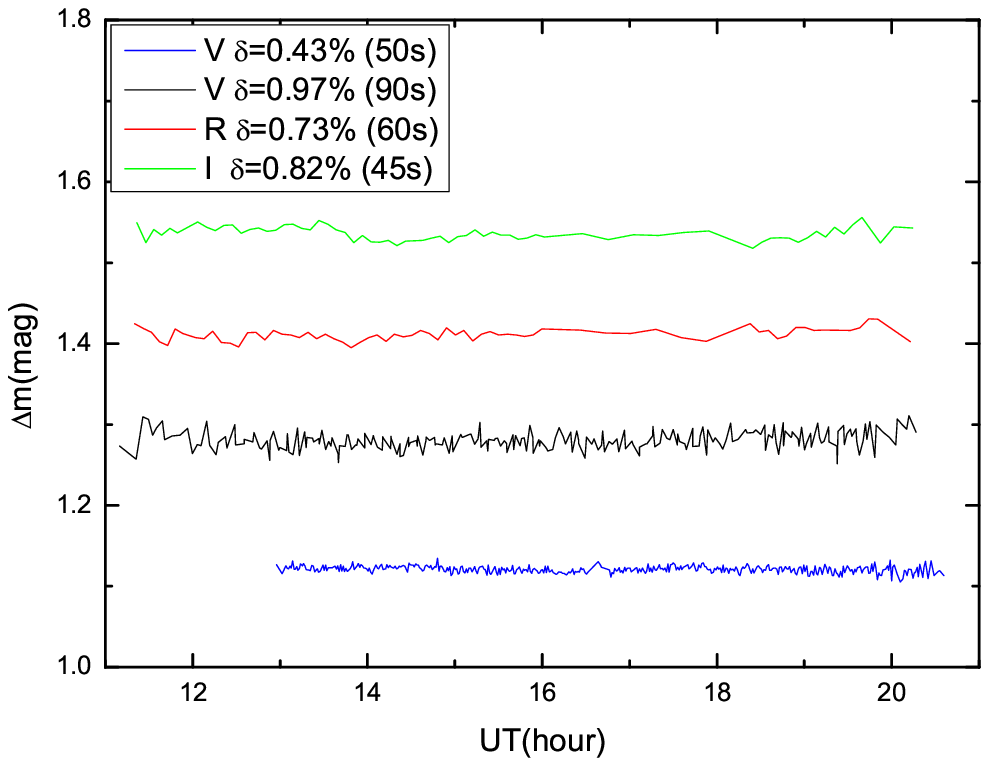}
    \end{minipage}
\end{tabular}
\caption{Left panel: The actual photometry accuracy of WHOT for \emph{V} band with 60\,s exposure. Right panel: Differential magnitude light curves for comparison stars in exoplanet transit HAT-P-33 field (blue line) and blazar BL Lac field (black, red and green line is for \emph{V}, \emph{R} and \emph{I} band, respectively).}
\label{accuracy}
\end{figure}

To estimate the internal photometry accuracy of WHOT, two 60-s exposures were done centered on NGC\,7790 in $V$ on 2010 October 9. The airmass of the two observations is 1.24 and 1.09, and the time interval is about two and half hours. Photometry of all sources with SNR higher than 2 were obtained and calibrated. The magnitude differences for the same source in two images were calculated, and plotted against the mean of two measurements, as shown in the left panel of Figure~\ref{accuracy}. This accuracy was under estimated especially for the fainter objects because the optics was dusty without cleaning for two years according to the working record.

Observations of blazars and transiting planets can be also used to estimate precision of differential CCD photometry. HAT-P-33 was monitored in $V$ for more than 7 hours during one of its planetary transit event on January 13, 2013, with exposure time of 50\,s. The differential light curve between two quiet reference stars ($V_1$=10.55 mag and $V_2$=10.72 mag ) in the HAT-P-33 FOV was shown in the right panel of Figure~\ref{accuracy} as blue solid line. The standard deviation of the light curve can be used to estimate the photometry precision \citep{zhou,mao}. The photometry precision is 0.0043 mag for \emph{V}$\approx$11 mag with 50\,s exposure from HAT-P-33 observations. Only \emph{V} filter was used for transit observation, so the same analysis was performed for two reference stars close to BL Lac in $V$, $R$ and $I$-bands. The reduced differential light curves between B and C \citep{smith} were plotted in the right panel of Figure~\ref{accuracy}. The photometry precision is 0.0097, 0.0073 and 0.0082 mag in \emph{V}($\approx$14.2 mag), \emph{R}($\approx$13.7 mag) and \emph{I}($\approx$13.2 mag) band, respectively. There are no such observations in \emph{U} and \emph{B} band, so we did not give the photometry precision in these two bands. The photometry precision depends on the brightness of target, exposure time and the observation conditions, but it is generally smaller than 0.01\,mag with appropriate setup under photometric conditions, based on our measurements.

\section{summary}
\label{sect:summary}

In order to let the astronomers know well and make good use of WHOT for their scientific observation research, we introduce the photometric system of WHOT, and evaluate its performance, and the site conditions of WHO in this paper. The image quality of the optic system is good and uniform over the detector. The detector has a very good linearity until a count of $\sim$50000, and has a low dark current due to a low working temperature. According to the observation log in the past 6 years, the seeing has a median value of $\sim$1.7\,arcsec, and is smaller than 2.0 arcsec in more than 85\% observation nights. No obvious seasonal change of seeing is found. The average number of clear nights at WHO is 155 per year, and the average observational time per year is 1126 hours. Summer season is the worst season, with only about 8.7 usable nights per month. The darkest night sky brightness at WHO is 20.99, 20.17, 18.90, 18.95 and 19.11 mag~arcsec$^{-2}$ in \emph{U}, \emph{B}, \emph{V}, \emph{R} and \emph{I} band, respectively, which are much brighter than those of other astronomical sites. Photometrical calibrations were performed using standard star observations taken in eight photometric nights. The average first-order atmosphere extinction coefficients are larger than those of Xinglong and Lulin observatories, due to its low elevation, while the color terms are consistent with those of Xinglong and Lulin. The peak system throughput is 10.9\%, 27.0\%, 38.8\%, 38.8\% and 31.8\% in \emph{U}, \emph{B}, \emph{V}, \emph{R} and \emph{I} band, respectively. The limiting magnitude with SNR of 100 and 300\,s exposure is 15.7, 16.7, 16.2, 16.1 and 15.9 mag for $U$, $B$, $V$, $R$ and $I$ band, respectively. The photometry precision is 4.3 mmag or better for $V\approx$11 mag with 50\,s exposure, and the photometry precision is 9.7, 7.3 and 8.2 mmag in \emph{V}($\approx$14.2 mag), \emph{R}($\approx$13.7 mag) and \emph{I}($\approx$13.2 mag) band, respectively.

\begin{acknowledgements}
We are grateful to the referee for insightful comments and valuable suggestions that have been adopted and consulted to improve this paper very much. We owe great thanks to all the staffs working at WHO who make great efforts to the telescope running and observations. We thank Profs. Gang Zhao and Xiaojun Jiang for their help for building and running WHO. This research is supported by the National Natural Science Foundation of China and Chinese Academic of Sciences joint fund on astronomy under project No. 10778701, 10778619 and by the National Natural Science Foundation of China under grants No. 11203016, 11333002, 11143012. This work is partly supported by the Natural Science Foundation of Shandong Province under grant No. ZR2012AQ008.

\end{acknowledgements}

\bibliographystyle{aa}

\begin{thebibliography}{}
\bibitem[\protect\citeauthoryear{Bessell} {1979}]{bessell1979} Bessell, M. S. 1979, PASP, 91, 589
\bibitem[\protect\citeauthoryear{Bessell} {2005}]{bessell} Bessell, M. S. 2005, ARA\&A, 43, 293
\bibitem[\protect\citeauthoryear{Bhatta et al.} {2013}]{bhatta} Bhatta, G., Webb J. R., Hollingsworth, H., et al. 2013, A\&A, 558, 92
\bibitem[\protect\citeauthoryear{Chen et al.} {2013}]{chen2013} Chen, X., Hu, S. M., \& Guo, D. F. 2013, arXiv:1311.5096
\bibitem[\protect\citeauthoryear{Chen et al.} {2014}]{chen2014} Chen, X., Hu, S. M., Guo, D. F., \& Du, J. J., 2014, Ap\&SS, 349, 909
\bibitem[\protect\citeauthoryear{Dai et al.} {2012}]{dai} Dai, H.-F., Yang, Y.-G., Hu, S.-M., \& Guo, D.-F. 2012, New Astronomy, 17, 347
\bibitem[\protect\citeauthoryear{Guo et al.} {2014}]{guo} Guo, D.-F., Hu, S.-M. \& Chen, X., JApA, 2014, DOI: 10.1007/s12036-014-9213-0
\bibitem[\protect\citeauthoryear{Howell} {2000}]{howell} Howell, S. B. 2000, Handbook of CCD Astronomy (ISBN 0-521-64834-3), Cambridge University Press

\bibitem[\protect\citeauthoryear{Huang et al.} {2012}]{huangfang} Huang, F., Li, J.-Z., Wang, X.-F., et al. 2012, RAA, 12, 1585

\bibitem[\protect\citeauthoryear{Hu et al.} {2014}]{hu2014} Hu, S. M., Chen, X.,\& Guo, D. F., 2014, JApA, DOI: 10.1007/s12036-014-9206-z

\bibitem[\protect\citeauthoryear{Kinoshita et al.} {2005}]{kinoshita} Kinoshita, D., Chen, C.-W., Lin, H.-C., et al. 2005, ChJAA, 5, 315

\bibitem[\protect\citeauthoryear{Landolt} {1992}]{landolt} Landolt, A. U. 1992, AJ, 104, 340
\bibitem[\protect\citeauthoryear{Li et al.} {2014}]{likai} Li, K., Hu, S.-M., Jiang, Y.-G., Chen, X. \& Ren, D.-Y., 2014, NewA, 30, 64

\bibitem[\protect\citeauthoryear{Liu et al.} {2010}]{liuli} Liu, L.-Y., Yao, Y.-Q., Wang, Y.-P., et al. 2010, RAA, 10, 1061

\bibitem[\protect\citeauthoryear{Liu et al.} {2003}]{liu} Liu, Y., Zhou, X., Sun, W.-H., et al., 2003, PASP, 115, 495

\bibitem[\protect\citeauthoryear{Mao et al.} {2013}]{mao} Mao, Y.-N., Lu, X.-M., Wang, J.-F., \& Jiang, X.-J. 2013, RAA, 13, 239

\bibitem[\protect\citeauthoryear{Smith et al.} {1985}]{smith} Smith, P. S., Balonek, T. J., Heckert, P. A., Elston, R., \& Schmidt, G. D. 1985, AJ, 90, 1184

\bibitem[\protect\citeauthoryear{Xu et al.} {2013}]{xudong} Xu, D., de Ugarte Postigo, A., Leloudas, G., et al. 2013, ApJ, 776, 98

\bibitem[\protect\citeauthoryear{Yao et al.} {2012}]{yao} Yao, S., Liu, C., Zhang, H.-T., et al. 2012, RAA, 12, 772

\bibitem[\protect\citeauthoryear{Yang et al.} {2010}]{yang} Yang, Y.-G., Hu, S.-M., Guo, D.-F., Wei, J.-Y., \& Dai, H.-F. 2010, AJ, 139, 1360

\bibitem[\protect\citeauthoryear{Zhang et al.} {2013}]{zhang} Zhang, H.-H., Liu, X.-W., Yuan, H.-B., et al. 2013, RAA, 13, 239

\bibitem[\protect\citeauthoryear{Zhou et al.} {2009}]{zhou} Zhou, A.-Y., Jiang, X.-J., Zhang, Y.-P., \& Wei, J.-Y. 2009, RAA, 9, 349


\bibitem[\protect\citeauthoryear{Zou et al.} {2010}]{zou} Zou, H., Zhou, X., Jiang, Z.-J., et al. 2010, AJ, 140, 602




\end{thebibliography}

\end{document}